\documentclass[10pt,twocolumn,letterpaper]{article}

\usepackage[pagenumbers]{cvpr}   

\usepackage{wrapfig}

\newcommand{\new}[1]{#1}
\newcommand{\newtag}{}
\newcommand{\revisedtag}{}
\newenvironment{newpar}{}{}

\definecolor{cvprblue}{rgb}{0.21,0.49,0.74}
\usepackage[pagebackref,breaklinks,colorlinks,allcolors=cvprblue]{hyperref}

\def\paperID{0}
\def\confName{Preprint}
\def\confYear{2026}

\title{EASE: Parametric Garment Design with Explicit and Local Ease Control}

\author{%
\parbox{\textwidth}{\centering
Kristijan Bartol\textsuperscript{1,4} \quad
Frieda Hentschel\textsuperscript{1} \quad
Nataliya Sadretdinova\textsuperscript{2} \quad
Benjamin Russig\textsuperscript{1}\\[4pt]
Melinos Averkiou\textsuperscript{3} \quad
Yordan Kyosev\textsuperscript{2} \quad
Stefan Gumhold\textsuperscript{1}\\[6pt]
\normalsize
\textsuperscript{1}TU Dresden, Faculty of Computer Science, Dresden, Germany\\
\textsuperscript{2}TU Dresden, Institute of Textile Machinery and High Performance Material Technology (ITM), Dresden, Germany\\
\textsuperscript{3}CYENS Centre of Excellence, Nicosia, Cyprus\\
\textsuperscript{4}University of Zagreb, Faculty of Electrical Engineering and Computing, Zagreb, Croatia\\[4pt]
{\tt\small kristijan.bartol@tu-dresden.de}%
}}

\begin{document}
\raggedbottom
\maketitle

\begin{abstract}
Garment fit and comfort depend critically on ease, the local allowance of excess material relative to the body. In existing design pipelines, ease is typically a byproduct of geometry or simulation rather than an independent design variable, making it difficult to specify, edit, transfer, or redistribute without re-running simulation or optimization. We propose a garment representation that embeds meshes directly on the surface of a parametric human body model and represents ease explicitly as spatially varying, anisotropic per-triangle scales. These scales act as primary design variables, decoupling the specification of material allowance from its physical deformation. Given a design specified by parametric and user-defined surface cuts together with local scale fields, we optimize sewing patterns that enforce the prescribed ease distribution while satisfying geometric and seam constraints. The representation enables three capabilities that are unavailable without explicit ease control: (1) direct specification and editing of local material allowance on the body surface; (2) intent-preserving transfer to new body shapes that reproduces the specified ease distribution without re-running simulation; and (3) intent-modifying pose adaptation that redistributes ease to relieve strain in high-stretch regions. We verify each of these experimentally: ease is closely retained after optimization, excessive strain is significantly mitigated for target poses, and the ease distribution is accurately transferred to target shapes. The approach is implemented as a virtual try-on framework, with physics-based cloth simulation used for final garment visualization. We will publicly release our framework and detailed documentation.\footnote{Source code: \url{https://github.com/kristijanbartol/ease}}
\end{abstract}

\section{Introduction}

Computational fashion is an interdisciplinary research area spanning anthropometry, apparel theory, geometry processing, and physics-based cloth simulation, with applications in computer vision \cite{adversarial-body-measurement, bedlam, clothwild}, virtual fashion \cite{grasshopper3d, seamly2d}, and physics-based cloth simulation \cite{pbd}. In traditional apparel theory, garment design is formulated in terms of sewing patterns, their fit on a given body shape, and grading rules that adapt patterns across sizes \cite{patternmaking-for-fashion-design}. Computational approaches aim to automate parts of this process by establishing correspondences between 2D sewing patterns and 3D garment surfaces, which are subsequently optimized to satisfy fit requirements such as body shape, material behavior, and intended function. Once a garment is designed for a reference body, it is commonly modified in two complementary ways. Design transfer adapts a garment to a new body shape while preserving the designer’s intent, often expressed in terms of local fit or ease \cite{digital-garment-alternation, learning-shared-shape-space, parametric-virtual-try-on, dress-anyone, garment-refitting-for-digital-characters, contour-craft}. Design adaptation, in contrast, intentionally modifies local design to accommodate new poses or activities, for example by reducing excessive strain under motion \cite{digital-garment-alternation, wolff, montes, computational-pattern-making, rule-free-pattern-adjustment}.

\begin{figure*}
  \includegraphics[width=\textwidth]{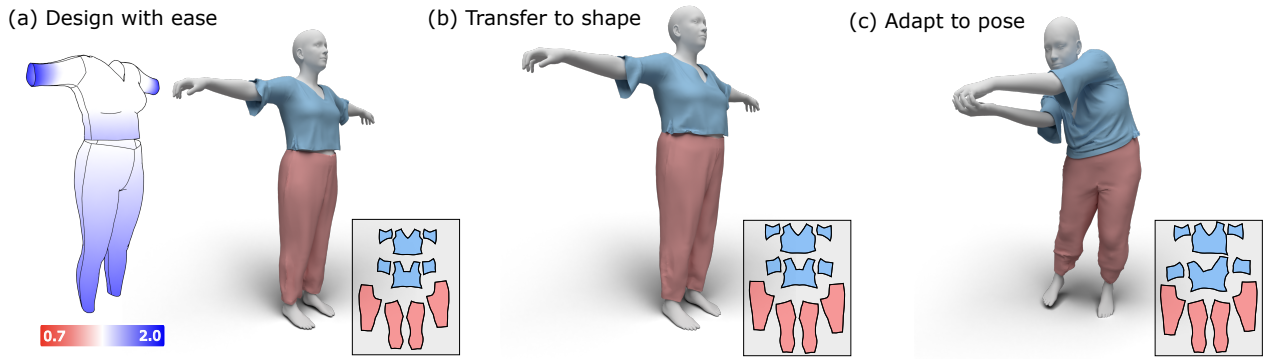}\vspace{-1.5em}
  \caption{Our framework supports three complementary tasks.
  (a)~\textit{Design with ease:} the designer paints a spatially varying ease field
  directly on the body surface (colormap: ease scale factor, where $1.0$ corresponds
  to zero material allowance; values above $1.0$ indicate progressively looser fit);
  the optimized sewing patterns (inset) are then draped via physics-based simulation.
  (b)~\textit{Transfer to shape:} the same design intent is transferred to a larger
  body shape while preserving the prescribed ease distribution.
  (c)~\textit{Adapt to pose:} the design is adapted to a target activity pose,
  redistributing ease to relieve strain in high-stretch regions.}
  \label{fig:teaser}
\end{figure*}

A key limitation of existing pipelines is that garment ease is not prescribed a priori as a spatially varying, anisotropic design input. Some approaches estimate a clothing gap or offset after simulation~\cite{meng2012-pattern, cordier2003, decaudin2006}, while others encode fit implicitly through stretch or strain fields derived from optimization~\cite{wolff, bartle, brouet}. Although effective for measuring or transferring fit, these representations encode ease only as a byproduct of geometry or simulation, not as a primary variable that drives the design process. As a consequence, editing, preserving, or redistributing ease requires re-running the corresponding geometric optimization or physical simulation. Parameterization-based methods~\cite{computational-pattern-making} recover sewing patterns from pre-designed 3D garment meshes. Ease in the resulting patterns is implicitly determined by the input geometry and cannot be edited independently. We address this limitation by embedding garment meshes directly on the surface of a parametric human body model \cite{smpl}, enabling a shared topology between body and garment, similar in spirit to \cite{garment-refitting-for-digital-characters, dress-anyone}. In this representation, ease is expressed directly as spatially varying, anisotropic local triangle scales, which act as design variables independent of physical deformation. This formulation enables optimization over design variables rather than deformed geometry or simulation states, supporting consistent transfer across body shapes and strain-free adaptation across poses.

To demonstrate this representation, we implement a virtual try-on pipeline in which users define garment topology through parametric and user-defined surface cuts and specify local ease via triangle scale fields. Sewing patterns are computed through constrained parameterization that recovers the prescribed ease distribution while satisfying geometric and seam constraints. To visualize the final designs and evaluate fit via body-garment gap measurements, we drape garments using extended position-based dynamics in Newton \cite{warp}, constrained by the optimized 2D sewing pattern. We summarize our contributions as follows:

\begin{itemize}
\item A garment representation that expresses spatially varying, anisotropic ease directly on a parametric human body surface, enabling localized and editable control of material allowance;
\item An optimization framework for intent-preserving design transfer across body shapes and intent-modifying design adaptation across poses;
\item A complete pipeline that produces sewing patterns consistent with a specified ease distribution, with physics-based cloth simulation used for final garment visualization.
\end{itemize}

\section{Related Work}

Our work relates to virtual garment modeling, optimization of sewing patterns to target shapes and poses, and cloth simulation. We focus on the garment design representation, and simulation is used as a downstream visualization tool.

\textbf{Garment modeling and design.} Professional 3D software \cite{clo3d} are the essential tools for creating original garment designs. However, design adaptation and transfer are typically manual and time consuming. Parametric design tools \cite{grasshopper3d, seamly2d} automate design adaptation but are based on 2D sewing patterns and \textit{grading}, which do not directly account for detailed 3D body shape variation. Wang et al.~\cite{wang-design-automation} extend this direction with automated pattern construction from body measurements, incorporating ease as fixed offsets derived from drafting rules. A seminal work by Umetani et al. \cite{umetani} proposes to design simultaneously and iteratively in 2D and 3D. Similarly to us, CLOTH3D \cite{cloth3d} displaces garment meshes from the SMPL body mesh; however, vertex-wise displacements do not represent ease or fabric constraints as controllable design parameters. A separate line of work focusing on procedural pattern generation \cite{korosteleva-dataset, garmentcode, garmentcodedata} proposes methods for simulating garment design from 2D sewing patterns parameterized by human body measurements. On top of these works, \cite{sewformer, dress-code} learn to interpolate and generate new designs.
A separate line of earlier work designs garments directly on a 3D body reference. Cordier et al.~\cite{cordier2003} propose a made-to-measure system in which garments are fitted to a 3D mannequin via collision response, with ease arising implicitly as the body-garment gap. Decaudin et al.~\cite{decaudin2006} sketch garment contours on a virtual mannequin, develop panels into sewing patterns, and encode a global ease through the body-surface offset, though this ease is not editable or anisotropic. Keckeisen et al.~\cite{keckeisen2004} enable interactive VR garment prototyping with automatic propagation of 3D edits to 2D patterns. Lu et al.~\cite{lu2017} apply the same 3D-to-2D paradigm to textile print design.
In contrast, our approach enables direct local control over ease in 3D.

\textbf{Optimization to target shapes and poses.} In principle, to adapt garment designs to different (especially larger) body shapes, one could drape an initial garment and gradually inflate it toward the target shape in a cloth simulation \cite{learning-shared-shape-space}. Brouet et al.~\cite{brouet} address the same goal geometrically, encoding design intent (including ease) as constraints during pattern grading from a source body to a target body. Meng et al.~\cite{meng2012-flexible} resize 3D garments to new body shapes via body-correspondence warping and optimize user-specified feature curves to locally preserve garment shape. In closely related work, Meng et al.~\cite{meng2012-pattern} propose a three-phase CAD framework in which 2D pattern pieces are cross-parameterized onto the body surface, a hybrid geometric-physical simulation restores the garment to its prescribed size, and the resulting clothing gap is used for fit evaluation and interactive 2D/3D pattern alteration, making it the first work to identify the clothing gap as an explicit design quantity. However, the gap emerges as a simulation output rather than being prescribed a priori, and is isotropic. Our approach inverts this relationship: the designer specifies an anisotropic ease field per triangle before any simulation, and sewing patterns are optimized to satisfy it.
A group of works adapts designs via differentiable cloth simulation \cite{garment-refitting-for-digital-characters, diffavatar, dress-anyone, parametric-virtual-try-on, contour-craft}, either by optimizing a control cage for 2D sewing patterns \cite{diffavatar, dress-anyone} or satisfying additional criteria such as multi-layer collision handling \cite{contour-craft}. Digital Garment Alteration \cite{digital-garment-alternation} adapts existing designs to target bodies accounting for anisotropic fabric properties. Wang \cite{rule-free-pattern-adjustment} proposes rule-free pattern adjustment under fit constraints. For pose adaptation, Wolff et al.~\cite{wolff} interpolate between poses using cloth simulation, simultaneously updating local triangle stretches across configurations for multiple target poses. Bartle et al.~\cite{bartle} recover 2D sewing patterns from user-specified 3D shapes via an iterative physics-based fixed-point scheme. Montes et al.~\cite{montes} propose a Lagrangian-on-Lagrangian mesh embedding for skintight garments, enabling controlled sliding over the body surface. Geometry-processing approaches \cite{computational-pattern-making} formulate adaptation as anisotropic parameterization by penalizing stretch along warp and weft directions. Qi and Igarashi~\cite{perfecttailor} update 2D sewing patterns to reflect user-specified 3D edits while preserving per-triangle 2D--3D scale relationships. These methods assume garment meshes are predesigned and treat ease as a byproduct of optimization; our framework instead exposes seamlines and a localized ease field as coupled design parameters within a single optimization loop.

\textbf{Neural garment fitting (draping).} Garment meshes are commonly represented as displacements from the SMPL body mesh, which is particularly suitable for learning-based methods focusing on capturing \cite{clothcap}, modeling \cite{smplicit, isp}, and estimation \cite{detailed-accurate-estimation, bcnet, clothwild, garment-recovery-with-priors} of garments in different poses and shapes. Displacement-based models typically learn separate models for each garment type.
Related approaches \cite{smplicit, isp} learn implicit functions for upper and lower garments that predict unsigned distance fields from body surface query points. These neural models are expressive enough for garment estimation from images \cite{clothwild, garment-recovery-with-priors}, but their latent representations do not guarantee by construction that fabric material is retained across poses and shapes. Another line of work focuses on learning garment shape spaces from sketches and 2D patterns \cite{learning-shared-shape-space, parametric-virtual-try-on, neural-tailor, sewformer}, but these methods do not automatically adapt designs to new poses. Neural Jacobian Fields~\cite{neural-jacobian-fields} learn local deformation Jacobians for human body rigging. Overall, neural-based approaches primarily target reconstruction or estimation rather than design-time control of material allowance.

\textbf{Learnable cloth simulations.} Learnable cloth simulations provide a data-driven alternative for predicting garment deformation across motion. Recent methods learn garment dynamics by predicting displacements \cite{pbns, snug} or by propagating deformation through graph neural networks \cite{hood, manifold-aware-transformers}, demonstrating robustness to complex motions and modest shape variation.
Grigorev et al. \cite{contour-craft} learn garment–garment collision handling for multi-layer garments and demonstrate automatic garment transfer across body shapes. While these methods accelerate forward dynamics, they do not treat material allowance as a controllable design signal.

In contrast to prior methods that infer, transfer, or preserve ease as a byproduct of geometric or physical optimization, we introduce ease as an independent, spatially distributed design variable that can be preserved or intentionally modified depending on the target task.
\section{Method}
\label{sec:method}

We propose a design pipeline that represents garments as locally scaled meshes
embedded on a parametric human body model and supports efficient adaptation
across body poses and shapes.
At a high level, the pipeline consists of three stages:
(1) parametric garment design on the body surface,
(2) pattern optimization via constrained parameterization, and
(3) physics-based draping for visualization.

\subsection{Parametric Garment Design -- Cutting the Body Surface}
\label{subsec:garment-design}

We define garment designs directly on the surface of the parametric SMPL body
model~\cite{smpl}.
SMPL represents the human body as a differentiable function $\mathcal{M}(\boldsymbol{\theta}, \boldsymbol{\beta})$, where $\boldsymbol{\theta} \in \mathbb{R}^{72}$ encodes joint rotations and $\boldsymbol{\beta} \in \mathbb{R}^{10}$ encodes shape coefficients. Its fixed-topology template mesh shares vertex connectivity across all poses and shapes, so garment patches cut from one body instance transfer consistently to any other via barycentric coordinates, including local design parameters such as ease (Sec.~\ref{subsec:optimization-algorithm}).
Unless stated otherwise, all designs are specified on the T-pose body
$\mathcal{M}(\boldsymbol{\theta}=0,\boldsymbol{\beta}=0)$ (zero pose and mean shape).

\textit{Design parameters.}
In our framework, a garment design is parameterized by three groups of variables:
(i) global length parameters controlling garment extent (e.g., sleeve or dress
length),
(ii) per-triangle weft scales specifying local ease, and
(iii) placement parameters that define seamline keypoints on the body surface.
Some parameters are shared across multiple patches (e.g., front and back panels
of a dress share the same vertical length).

\begin{figure}
    \centering
    \includegraphics[width=1.0\linewidth]{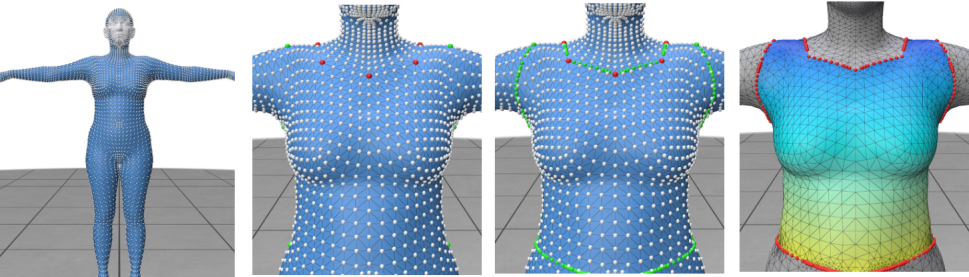}\vspace{-0.5em}
    \caption{The interactive user interface for keypoint selection and patch extraction. The keypoints are defined directly on the surface of the body mesh. Their order is used to determine the paths along which the mesh is cut. The cuts define seamlines and patch boundaries. Each local triangle has the corresponding local scale assigned, as indicated by the colors (right).}
    \label{fig:patch-extraction}
\end{figure}

\begin{figure}
\centering
    \includegraphics[width=0.6\linewidth]{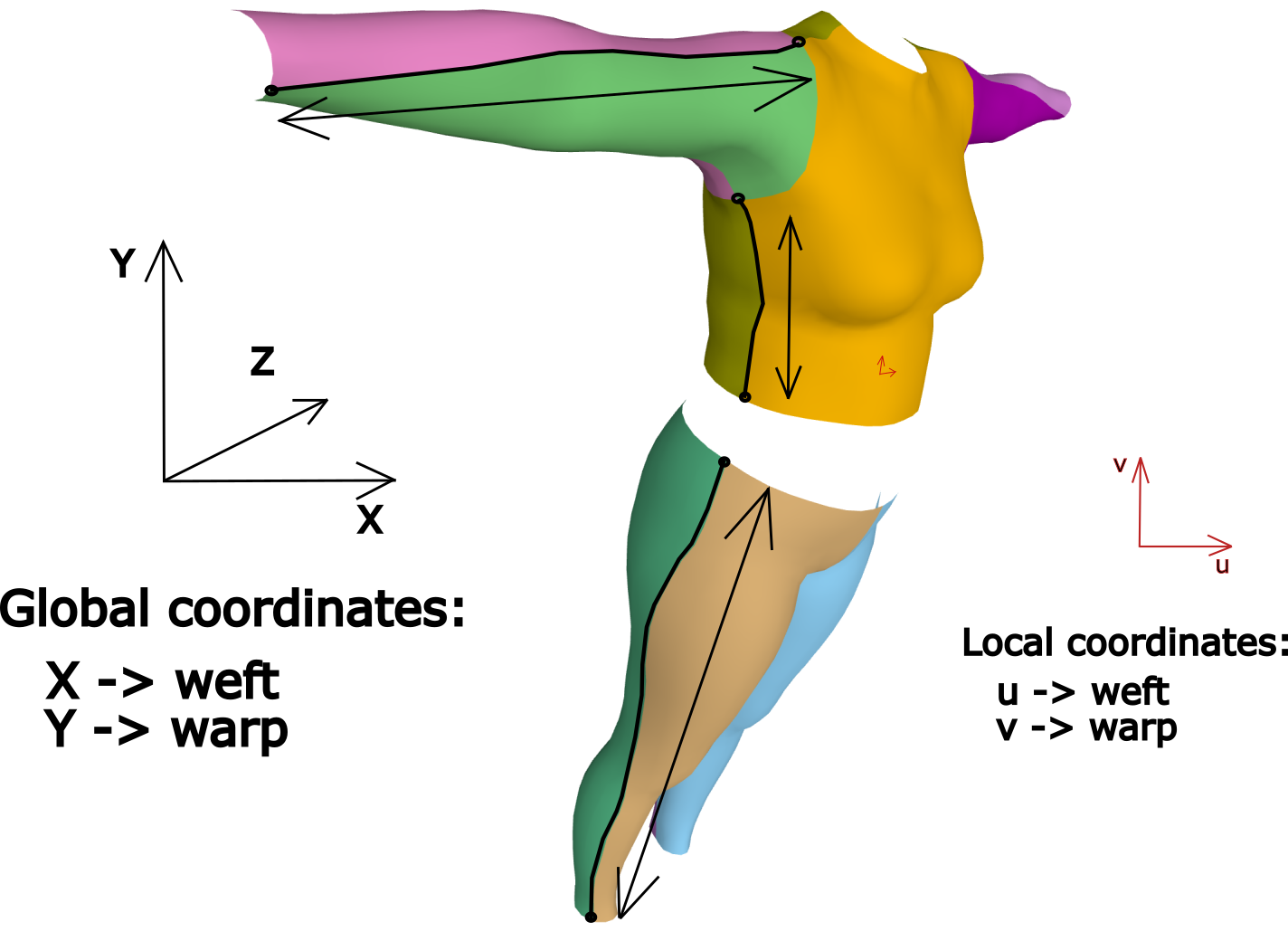}\vspace{-1.0em}
    \caption{Garment patches defined by seamlines on the body surface.}
    \label{fig:embedded-design}
\end{figure}

\textit{Seamlines and cutting.}
Seamline keypoints are defined on the SMPL body surface (Fig.~\ref{fig:patch-extraction}).
The keypoints include standard anatomical locations such as collar, shoulder, armpit,
and midline for upper garments, and waist, hip, and leg openings for lower-body
garments. Placement parameters specify positions along predefined body segments, while
length parameters define boundary points by tracing geodesic paths in
pose-dependent directions.
Symmetric counterparts are obtained by reflection across the body midline.
Keypoints are connected by geodesic paths to form closed seamlines, which are
inserted into the mesh and retriangulated to produce valid garment topology.
Each closed seamline defines a garment patch, which is extracted via flood fill
within the seam boundaries (Fig.~\ref{fig:patch-extraction}, \ref{fig:embedded-design}).
Topology changes introduced by cutting are transferred to target bodies using
barycentric coordinates.

\textit{Design templates.}
The full set of design parameters defines a garment template.
Templates can be predefined or interactively modified by adjusting parameters.

\textit{Reference and target bodies.}
We specify a reference body
$\mathcal{M}(\boldsymbol{\theta}_{\text{ref}},\boldsymbol{\beta}_{\text{ref}})$,
a set of target poses
$\mathcal{T}=\{\boldsymbol{\theta}_{\text{target}}^1,\ldots,\boldsymbol{\theta}_{\text{target}}^N\}$,
and optional target shapes
$\mathcal{S}=\{\boldsymbol{\beta}_{\text{target}}^1,\ldots,\boldsymbol{\beta}_{\text{target}}^M\}$,
which together define the body meshes used for adaptation and transfer.

\textit{Parametric ease.}
Each triangle of a garment patch, embedded on the body surface, is assigned a scalar
$\rho_{\text{weft}}^i$ that specifies the desired local ease of the garment
relative to the body surface.
Values $\rho_{\text{weft}}>1$ correspond to increased material allowance (looser
fit), while values below one produce tighter garments.
Unlike approaches that measure ease as the body-garment gap after simulation~\cite{meng2012-pattern, cordier2003}, we treat $\rho_{\text{weft}}$ as a primary design variable specified before any optimization or simulation.
This allows ease to be directly edited, preserved across body shapes, or redistributed across poses within a single optimization loop, without requiring re-simulation.
Warp-direction scales are fixed to $\rho_{\text{warp}}=1.0$ at the design stage,
as garment length is controlled explicitly through global length parameters. This separates local ease (weft) from length (global parameters); warp may exceed 1.0 only under pose adaptation (Sec.~\ref{subsec:optimization-algorithm}), where the target body requires it.
Fabric grain directions (weft and warp) are aligned to the global body axes: weft along the horizontal ($x$) axis and warp along the vertical ($y$) axis.
Since LSCM parameterization does not guarantee this alignment, we apply a global rotation to the 2D patch at each iteration to align it with the body axes ($x$ for weft, $y$ for warp), using an area-weighted vote over per-triangle alignment angles.\footnote{Near-degenerate edges (length below a numerical threshold) default to a fixed tangent direction.}
The effect of these scales on the resulting sewing pattern is defined during
pattern optimization (Sec.~\ref{subsec:optimization-algorithm}).

\begin{figure}
    \centering
    \includegraphics[width=1.0\linewidth]{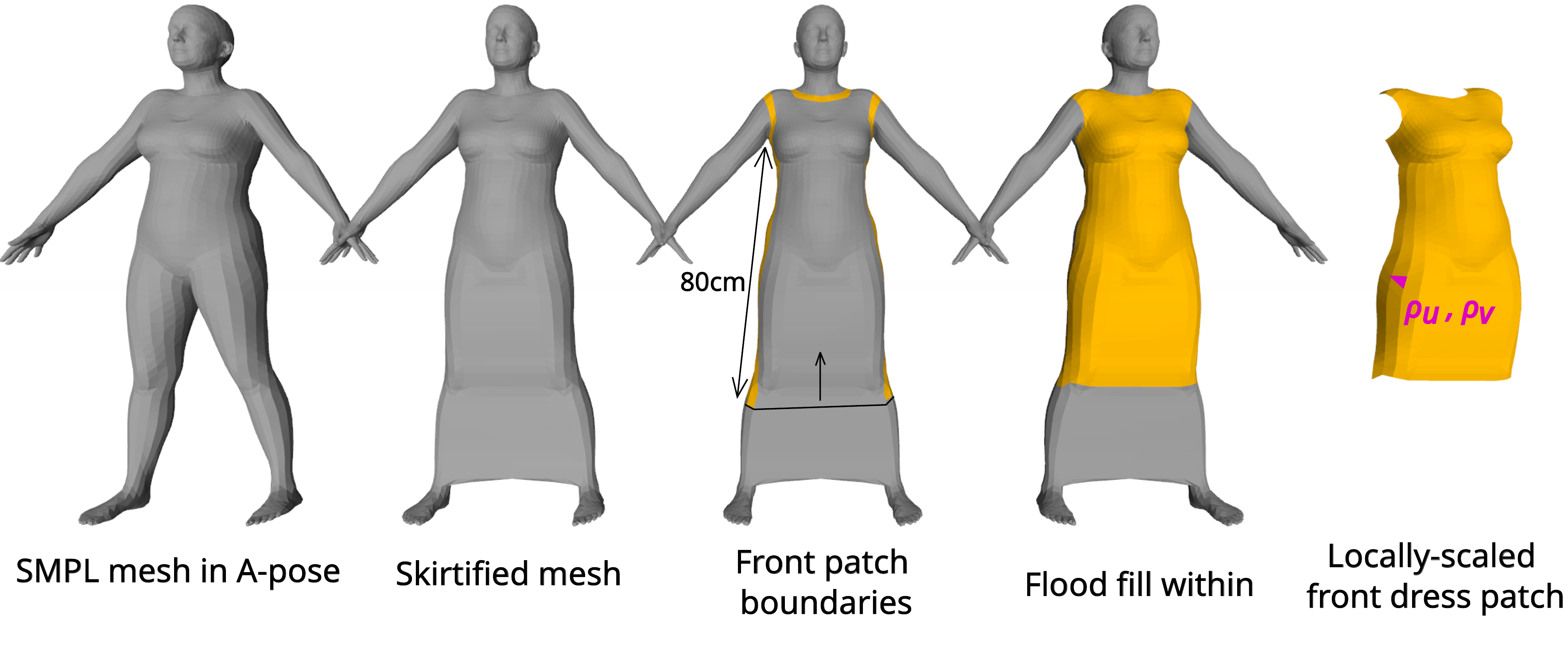}
    \caption{Extracting embedded 3D patches (front) directly on the parametric mesh. For dresses and skirts, we use a "skirtified" mesh topology.}
    \label{fig:design-process}
\end{figure}

\textit{Dresses and skirts.}
To model dresses and skirts, we modify the body topology prior to cutting.
We remove a set of faces in the inter-leg region of the SMPL mesh and connect the two exposed leg boundaries into a single closed boundary loop, producing a unified lower-body surface with skirt-like topology, similar to~\cite{cloth3d} (Fig.~\ref{fig:design-process}).

\subsection{Parameterization -- Optimizing Ease}
\label{subsec:optimization-algorithm}


For each garment patch $p$, we are given an embedded 3D triangle mesh
$S^p=(\mathcal{V}^p,\mathcal{F}^p)$ defined on the body surface, together with seam
correspondences to adjacent patches and a local fabric grain directions
per triangle.
Our goal is to compute a 2D sewing pattern, represented by 2D vertex positions
$\mathbf{u}^p=\{\mathbf{u}_v\in\mathbb{R}^2\}_{v\in\mathcal{V}^p}$.
We denote the optimized solution by $\mathcal{P}_*$.

The optimization minimizes an energy of the form
\begin{equation}
E =
\lambda_{\text{ease}} (E_{\text{weft}} + E_{\text{warp}})
+ E_{\text{seam}}
+ \lambda_{\text{reg}} E_{\text{reg}},
\end{equation}
where $\lambda_{\text{ease}}$ and $\lambda_{\text{reg}}$ are scalar weights. Here, $E_{\text{weft}}$ and $E_{\text{warp}}$ enforce anisotropic stretch
constraints along the fabric grain, $E_{\text{seam}}$ couples adjacent pattern
pieces along seams, and $E_{\text{reg}}$ penalizes deviation from isometric transformations per triangle, implicitly suppressing shear in the parameterization following the ARAP-based formulation of~\cite{arap}. We initialize the optimization using an LSCM
parameterization~\cite{lscm}, following standard practice~\cite{montes, computational-pattern-making, chebyshev-parameterization}. The optimization is performed independently for each patch, with seams coupling adjacent patches through shared boundary constraints.

\paragraph{Stretch constraints with per-triangle scales}
For a triangle with 2D vertex coordinates
$\mathbf{U}=[\mathbf{u}_A,\mathbf{u}_B,\mathbf{u}_C]\in\mathbb{R}^{2\times 3}$ and corresponding 3D coordinates
$\mathbf{X}=[\mathbf{x}_A,\mathbf{x}_B,\mathbf{x}_C]\in\mathbb{R}^{3\times 3}$, a point with barycentric weights
$\mathbf{b}\in\mathbb{R}^3$ is evaluated as $\mathbf{U}\mathbf{b}$ in 2D and
$\mathbf{X}\mathbf{b}$ in 3D.
Let $\mathbf{b}_c$ denote the barycentric coordinates of the triangle
centroid, and $\mathbf{b}_{\text{weft}}$ those of a unit offset from the
centroid in the weft direction of the 2D parameter domain.
The barycentric coordinates $\mathbf{b}_{\text{weft}}$ are computed once from the LSCM initialization and held fixed throughout all subsequent iterations, making the stretch constraints linear in the 2D vertex positions $\mathbf{U}$.
The corresponding 2D and 3D weft offsets are

\begin{equation}
\label{eq:pietroni-stretch-eq}
\boldsymbol{\sigma}_{2D} = \mathbf{U}(\mathbf{b}_{\text{weft}}-\mathbf{b}_c),
\qquad
\boldsymbol{\sigma}_{3D} = \mathbf{X}(\mathbf{b}_{\text{weft}}-\mathbf{b}_c).
\end{equation}

Following~\cite{computational-pattern-making}, \new{local stretch is minimized when} $\|\boldsymbol{\sigma}_{2D}\| = \|\boldsymbol{\sigma}_{3D}\|$ \new{along the fabric grain.}
We generalize this constraint by introducing a per-triangle weft scale
$\rho_{\text{weft}}$, which directly encodes the desired local ease.
For a reference body $\mathcal{M}(\boldsymbol{\theta}_{\text{ref}},\boldsymbol{\beta}_{\text{ref}})$, the
target 2D weft offset is defined as
\begin{equation}
\label{eq:ref-scale}
\boldsymbol{\sigma}_{2D}^{*} = \rho_{\text{weft}}\,\boldsymbol{\sigma}_{3D}^{\text{ref}}.
\end{equation}
The warp direction enforces unit scale ($\rho_{\text{warp}}=1.0$), i.e., $\boldsymbol{\sigma}_{2D}^{*,\text{warp}} = \boldsymbol{\sigma}_{3D}^{\text{warp,ref}}$, consistent with garment length being controlled through global length parameters. During pose adaptation, the warp scale may increase beyond 1.0 via the max formulation (Eq.~\ref{eq:adapt-scale}), overriding the design-time unit scale where the target pose requires it.

Both the weft and warp targets are enforced as least-squares terms and define the reference
pattern $\mathcal{P}_R$, in which the prescribed scale field explicitly represents
the intended ease distribution.

\begin{figure}
    \centering
    \includegraphics[width=1.0\linewidth]{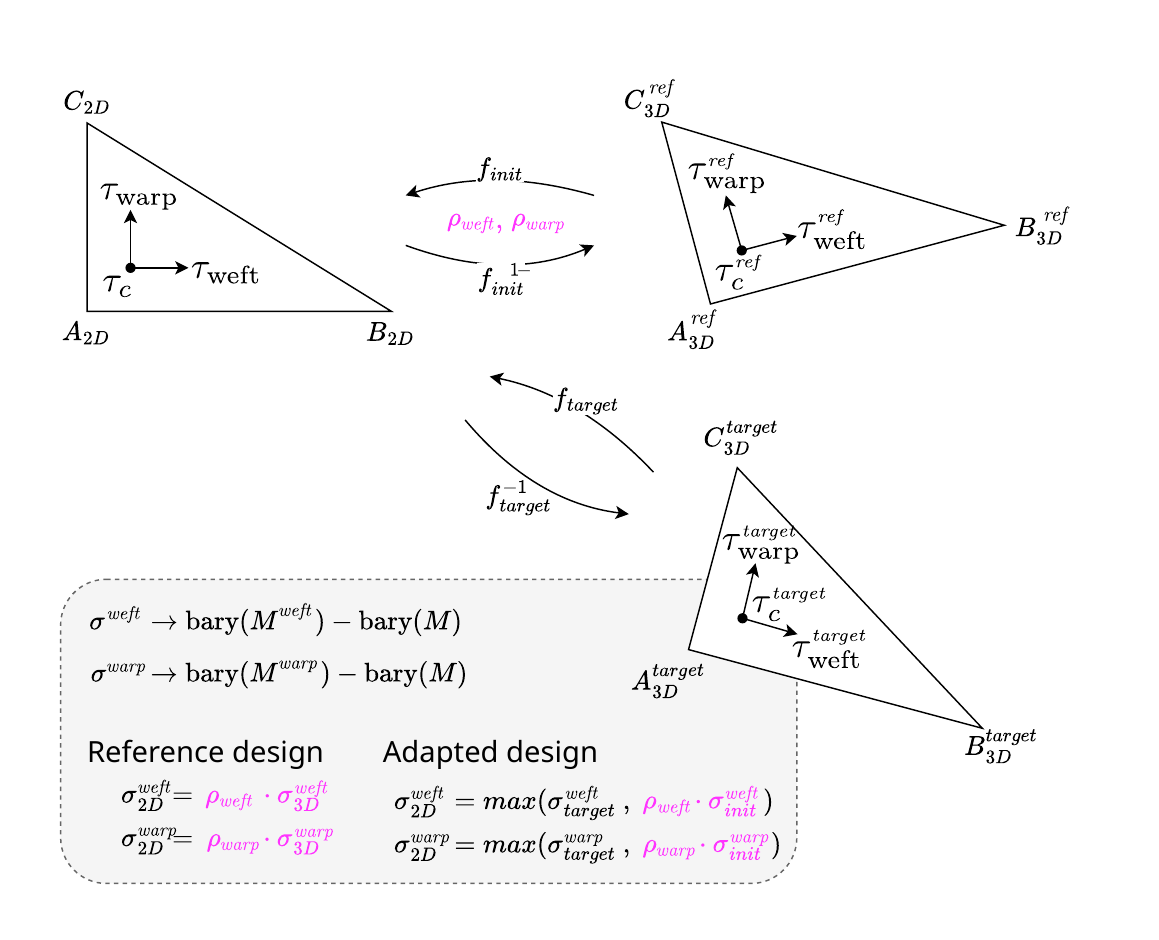}
    \caption{The illustration of the transformations between the 2D and 3D triangles of the reference and the target meshes with the corresponding stretch equations for pose adaptation.}
    \label{fig:stretch-equations}
\end{figure}

\paragraph{Adaptation, transfer, and material tolerance}
The same formulation naturally supports both pose adaptation and shape transfer
by modifying the target offset $\boldsymbol{\sigma}_{2D}^{*}$.
For adaptation to a set of target poses
$\{\boldsymbol{\theta}_{\text{target}}^i\}$, we require the pattern to be sufficiently large to cover the body in each target pose while preserving the prescribed ease where possible:
\begin{equation}
\label{eq:adapt-scale}
\boldsymbol{\sigma}_{2D}^{*} =
\max\!\left(
\rho_{\text{weft}}\,\boldsymbol{\sigma}_{3D}^{\text{ref}},
\;
\max_i \boldsymbol{\sigma}_{3D}^{\text{target},i}
\right),
\end{equation}
where the $\max$ is applied independently in each grain direction ($\rho_{\text{warp}}=1.0$ serves as the warp reference scale), and $\boldsymbol{\sigma}_{3D}^{\text{target},i}$ is evaluated on the deformed body mesh in pose $\boldsymbol{\theta}_{\text{target}}^i$ (Fig.~\ref{fig:stretch-equations}). \new{This formulation produces a single pattern that fits every target pose. The maximum is computed independently for the weft and warp directions, so the pattern accommodates the largest weft demand and the largest warp demand across all target poses.}
For transfer to a target body shape $\boldsymbol{\beta}_{\text{target}}$, we replace the
reference geometry by the target shape and define
\begin{equation}
\label{eq:transfer-scale}
\boldsymbol{\sigma}_{2D}^{*} =
\rho_{\text{weft}}\,\boldsymbol{\sigma}_{3D}^{\text{target}},
\end{equation}
which corresponds to grading the pattern while preserving the local ease field.

\new{The conservative pattern from Eq.}~\ref{eq:adapt-scale} \new{can leave surplus fabric in the reference pose. We mitigate this surplus by introducing a stretch tolerance} $\epsilon_{\text{weft}}>1$\new{, modifying Eq.}~\ref{eq:adapt-scale} \new{as}
\begin{equation}
\label{eq:material-scale}
\boldsymbol{\sigma}_{2D}^{*} =
\max\!\left(
\rho_{\text{weft}}\,\boldsymbol{\sigma}_{3D}^{\text{ref}},
\;
\frac{\max_i \boldsymbol{\sigma}_{3D}^{\text{target},i}}{\epsilon_{\text{weft}}}
\right),
\end{equation}
allowing controlled relaxation of the constraints when permitted by the fabric.
The same tolerance applies symmetrically to the warp direction as $\epsilon_{\text{warp}}$.
All cases are handled within the same iterative local-global optimization.

\textit{Seam constraints.}
Each sewing seam is represented by a pair of corresponding boundary polylines on
two 2D pattern pieces, say patches $p$ and $q$.
We denote the ordered vertex sequences along the seam as
$\Gamma^{p\rightarrow q}=\{\mathbf{u}^{p}_1,\ldots,\mathbf{u}^{p}_N\}$ and
$\Gamma^{q\rightarrow p}=\{\mathbf{u}^{q}_1,\ldots,\mathbf{u}^{q}_N\}$, where
$\mathbf{u}^{p}_i,\mathbf{u}^{q}_i\in\mathbb{R}^2$ are corresponding boundary vertices on the two
patterns.

The seam alignment energy $E_{\text{seam}}$ couples adjacent pattern pieces by encouraging compatible seam geometry. It consists of three weighted sub-terms,
\begin{equation}
\label{eq:seam-energy}
E_{\text{seam}} = \lambda_{\text{len}} E_{\text{len}} + \lambda_{\text{straight}} E_{\text{straight}} + \lambda_{\text{curve}} E_{\text{curv}},
\end{equation}
corresponding to length compatibility, straightness, and curvature alignment, respectively.
To maintain a tractable optimization, all seam terms are defined per corresponding edge or vertex.

\paragraph{Length compatibility}
Following prior work on sewable pattern generation (e.g., \cite{montes}), we
enforce compatibility of corresponding seam lengths.
Let $\mathbf{e}^{p}_i=\mathbf{u}^{p}_{i+1}-\mathbf{u}^{p}_i$ and $\mathbf{e}^{q}_i=\mathbf{u}^{q}_{i+1}-\mathbf{u}^{q}_i$ denote
corresponding seam edges.
We add the per-edge term
\begin{equation}
E_{\text{len}} =
\sum_{i=1}^{N-1}
\left(\lVert \mathbf{e}^{p}_i\rVert - \lVert \mathbf{e}^{q}_i\rVert\right)^2 ,
\label{eq:seam-length}
\end{equation}
which encourages the two seamlines to have matching arc length and improves
fabricability.

\paragraph{Seam regularization}
In addition, we include regularization terms to improve seam
quality. To discourage high-frequency oscillations, we penalize the discrete second
derivative of each seam polyline,
\begin{equation}
E_{\text{straight}} =
\sum_{i=2}^{N-1}
\big\lVert \mathbf{u}^{p}_{i+1}-2\mathbf{u}^{p}_i+\mathbf{u}^{p}_{i-1} \big\rVert^2
+
\sum_{i=2}^{N-1}
\big\lVert \mathbf{u}^{q}_{i+1}-2\mathbf{u}^{q}_i+\mathbf{u}^{q}_{i-1} \big\rVert^2 ,
\label{eq:seam-straight}
\end{equation}
which promotes locally smooth seamlines.
Although the two sides of each seam coincide on the body, the parameterization may yield different 2D shapes on either side, producing seamlines that are individually smooth but locally non-matching. We address this by matching the discrete second derivatives of the corresponding 2D polylines, encouraging shared convex or concave segments:

\begin{equation}
E_{\text{curv}} =
\sum_{i=2}^{N-1}
\big\lVert
(\mathbf{u}^{p}_{i+1}-2\mathbf{u}^{p}_i+\mathbf{u}^{p}_{i-1})
-
(\mathbf{u}^{q}_{i+1}-2\mathbf{u}^{q}_i+\mathbf{u}^{q}_{i-1})
\big\rVert^2 .
\label{eq:seam-curv}
\end{equation}

\begin{figure*}[!t]
    \centering
    \includegraphics[width=0.8\linewidth]{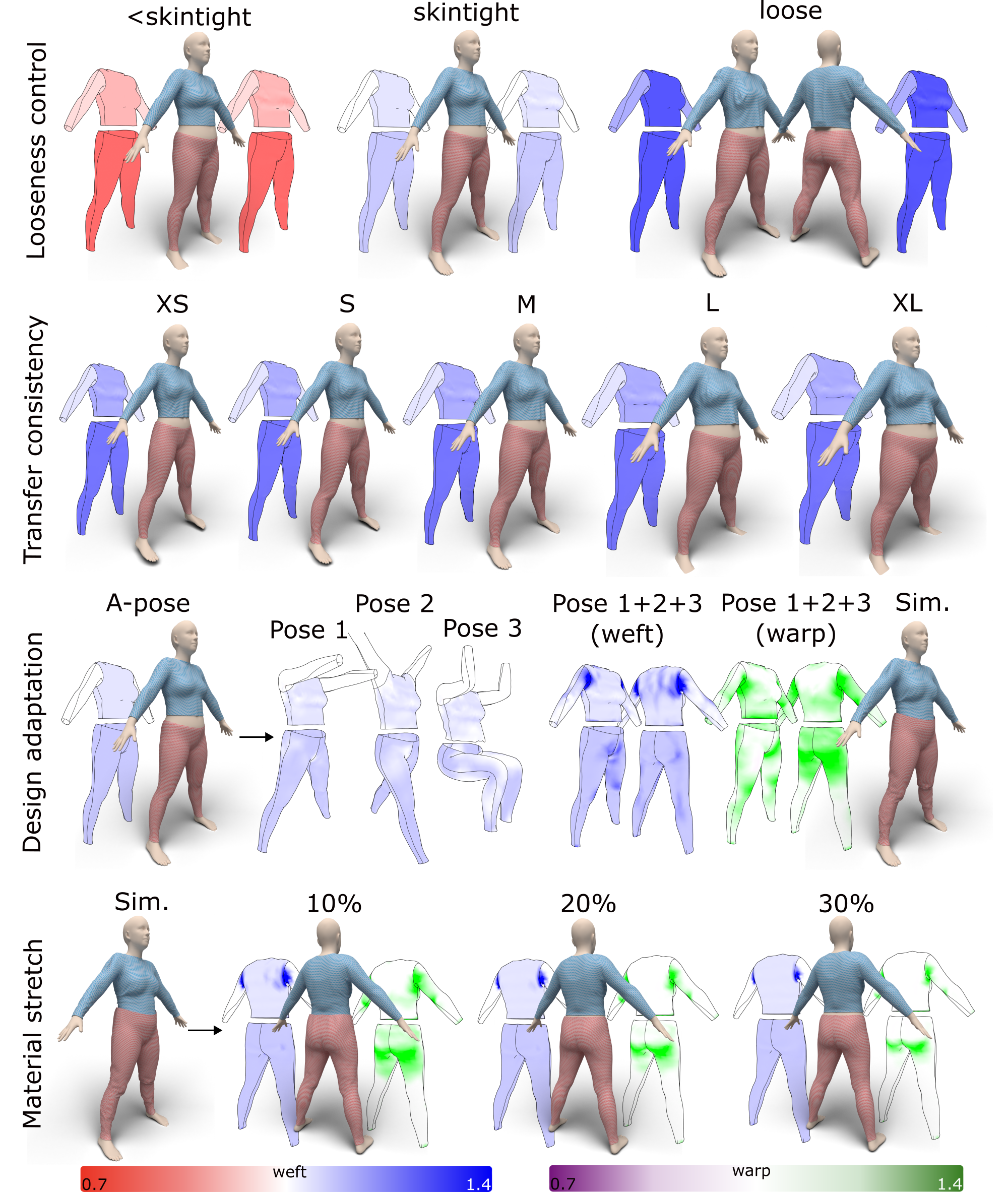}
    \vspace{-1em}
    \caption{Qualitative evaluation of ease. In the first row, the specified ease (left) corresponds to the simulated garment (middle) and the optimized ease distribution (right). The second row shows transfer consistency between five body shapes. The third row  adapts to three target poses independently (left), and jointly (right). Both weft and warp scales are shown. Finally, the material stretch tolerance is specified to avoid excessive fabric in the reference pose.}
    \label{fig:qualitative}
\end{figure*}

In practice, we use all seam terms jointly with fixed relative weights across all experiments; we found this combination to produce stable, sewable patterns without manual tuning.

\textit{Optimization solve.}
The optimization is solved via an iterative local-global scheme~\cite{arap}. In each local step, per-triangle rotations and seam alignment targets are updated based on the current solution; the seam length term $E_{\text{len}}$ is linearized per iteration by projecting onto the current edge tangent direction, with target lengths set as the average of the two sides' current edge lengths. In the global step, a sparse linear least-squares system is solved for the 2D vertex positions with these targets fixed. We run 20 iterations, after which all solutions are empirically observed to converge. We first solve for the reference body to obtain the reference pattern $\mathcal{P}_R$; for pose adaptation or shape transfer, the same procedure is applied with updated right-hand-side terms reflecting the target body geometry, while keeping the seam pairing fixed. The final optimized pattern is denoted by $\mathcal{P}_*$. From $\mathcal{P}_*$, we extract per-triangle fitted scale factors $\Pi_*^{p,i}$ using the same barycentric constructions as in Eqs.~\ref{eq:pietroni-stretch-eq}–\ref{eq:ref-scale}, which are later used for transfer and simulation (Sec.~\ref{subsec:simulation-stage}).

\subsection{Garment Draping - Cloth Simulation}
\label{subsec:simulation-stage}

To visualize the resulting garment designs, we generate draped 3D garments using physics-based cloth simulation. Garment patches are first grouped into upper and lower garments and stitched by merging corresponding patch meshes and removing duplicate seam vertices, producing skintight reference meshes $(\boldsymbol{x}^{\text{upper}}, \boldsymbol{x}^{\text{lower}})$ embedded on the body surface. The optimized sewing patterns, $\mathcal{P}_*$ obtained from the parameterization stage, define correspondences between 2D (pattern) and 3D (embedded) triangles. We use this correspondence to initialize a cloth simulation based on Newton’s XPBD solver, which takes the 2D sewing pattern, seam connectivity, and material parameters as input. From the optimized pattern, the solver derives target rest lengths and anisotropic stretch constraints in the weft and warp directions, which are used to initialize the cloth in a non-skintight configuration. Cloth dynamics, gravity, and collisions with the body are resolved using XPBD constraints. To maintain stable attachment and prevent global drift, selected boundary vertices (e.g., collar for upper garments and waistline for lower garments) are pinned during simulation.
\section{Evaluation}
\label{sec:evaluation}

In this section, we verify that the framework can (Fig.~\ref{fig:qualitative}):

\begin{itemize}
    \item specify design ease as an explicit variable that is optimized with other geometry constraints to the reference mesh (first row),

    \item consistently optimize transferred design to target body shapes (second row),

    \item adapt the reference or transferred design to one or more target body poses (third row).
\end{itemize}

\noindent We additionally validate material stretch control via the tolerance factors $\epsilon_{\text{weft}}$, $\epsilon_{\text{warp}}$ (last row in Fig.~\ref{fig:qualitative}), drape the optimized designs using cloth simulation (all rows), and verify that the prescribed ease faithfully translates to physical body-garment gap after draping (Fig.~\ref{fig:gap-graph}).

As a real-world verification of the patterns, we manufacture a piece of lower garment for the male subject (Fig.~\ref{fig:pant-manufacturing}).

\subsection{Qualitative evaluation}

\textbf{Ease control $\Pi_{p_i}$.} The first row highlights different uniform looseness values for sleeves, torso, and pants, respectively, for the average body shape $\boldsymbol{\beta}_{\text{avg}}$. The first group shows tightwear $(0.95, 0.9, 0.8)$
, the second shows regular fit $(1.0, 1.05, 1.1)$, and the third shows casual fit $(1.1, 1.3, 1.3)$. In addition, the first sample in each group shows the specified scalars $\Pi$, and the third shows the optimized scalars $\Pi_*$ with only subtle differences.

\textbf{Design transfer to target shapes.} The second row shows five different body shapes, labeled from \textit{XS} to \textit{XL}, as is common in the fashion industry\footnote{The garment sizes do not correspond to any standard.}, for uniform ease values $(1.05, 1.1, 1.2)$. We highlight the optimized ease distribution and the corresponding simulation for each size. We visually verify that the design remains consistent across various shapes both in terms of ease distribution and appearance (geometry and wrinkle details).

\textbf{Design adaptation to target poses.} The third row shows examples of designs adapted from the reference A pose to the target poses $\mathcal{T} = \{\boldsymbol{\theta}_1, \boldsymbol{\theta}_2, \boldsymbol{\theta}_3\}$, for the average body shape. The first example adapts to the arms-front pose with spread legs (abv. P1). The second example adapts to the arms-up pose with scissored legs (abv. P2). The third example adapts to the sit-pose with bent arms in front of the body (abv. P3). On the right side, the initial design is adapted to all three poses. The optimized scalars $\Pi_*$ are visualized w.r.t. the initial A-pose, both in the weft (blue) and warp (green) directions. It can be observed that significant local stretches have appeared along both orthogonal directions, which is also reflected in the visibly looser draped garment. The critical areas appear around the shoulder area along the weft direction and around the back arms and back upper legs along the warp direction. The "stretches", i.e., the optimized scales along the warp (vertical) direction, result in the elongation of the garments, as expected.

\textbf{Material stretch control.} The last example ("Pose 1+2+3"), jointly adapted from A-pose to P1, P2, and P3, produces garments with a significant amount of excessive fabric, which makes the garment less suitable for wearing in the reference, A-pose. Therefore, in the last row in Fig.~\ref{fig:qualitative}, we vary the anisotropic material stretch factors $\epsilon_{\text{weft}}$ and $\epsilon_{\text{warp}}$\footnote{For simplicity of the analysis and presentation $\epsilon_{\text{weft}} = \epsilon_{\text{warp}}$.} at 10\%, 20\%, and 30\%, corresponding to the three groups of meshes. The corresponding scale values for the weft (blue) and warp (green) direction drop with the increase of the material stretch factors, as expected.

\subsection{Quantitative evaluation}

We quantitatively evaluate compliance with the specified local ease distribution and the consistency of ease under design transfer across body shapes. Ease compliance is measured as the normalized deviation between the design-specified and achieved weft scale per triangle (deviation divided by specified scale), evaluated across a representative top and a pair of pants for uniform ease values between 0.9 and 2.0. Table~\ref{tab:ease-compliance} reports results aggregated across both garments for three ease ranges. Compliance degrades moderately from approximately 1.3\% at tight ease to 2.2\% at loose ease, reflecting a known competition between the ease term and the regularization term, which anchors edge lengths to their original 3D values. All values remain below 2.5\% mean normalized error across the full tested range, indicating that the specified ease is accurately realized for both garment types.

To evaluate design transfer, we measure the per-triangle deviation in local ease after transferring the initial design to four target body shapes (XS, S, L, XL). Deviations are normalized by the specified scale to remain comparable across ease levels. As shown in Table~\ref{tab:transfer}, ease is preserved within 0.6--1.0\% for moderate shape changes (S, L) and within 1.8\% for the most extreme shape change (XL).

To evaluate design adaptation, we measure the percentage of garment surface area that requires stretch beyond the material tolerance, before and after adaptation. This is computed as the fraction of total garment area where the per-triangle weft stretch exceeds $\epsilon_{\text{weft}}$. Table~\ref{tab:residual-stretch} reports results for poses P1, P2, and P3 for both upper and lower garments. For the upper garment, 27--34\% of the surface area is stretched before adaptation, reduced to 4.3--4.5\% after. The lower garment is less affected by the tested poses (P1 and P2 primarily involve arm movement), but still shows consistent reduction, with the sit pose (P3) producing the largest improvement (9.9\% to 2.5\%). Table~\ref{tab:seam-runtime} reports seam compatibility and runtime for both garment types.

\begin{newpar}
To verify that ease compliance generalizes beyond uniform fields, we evaluate two representative non-uniform designs already shown qualitatively in the paper: the simple dress with a linearly increasing top-to-bottom looseness (Fig.~\ref{fig:analyses}, first column), and the teaser shirt and pants (Fig.~\ref{fig:teaser}), which combines per-component uniform values across the front, back, and sleeves. These cover the two main families of non-uniform specification a designer is likely to author: smooth gradients within a single component, and piecewise-uniform values across multiple components. The bottom rows of Table~\ref{tab:ease-compliance} report per-triangle compliance for each design, remaining within the same range as the uniform case and indicating that the optimization handles spatially varying targets without systematic degradation.
\end{newpar}

\subsection{Post-drape gap evaluation}

\begin{figure}[t]
    \centering
    \includegraphics[width=0.85\columnwidth]{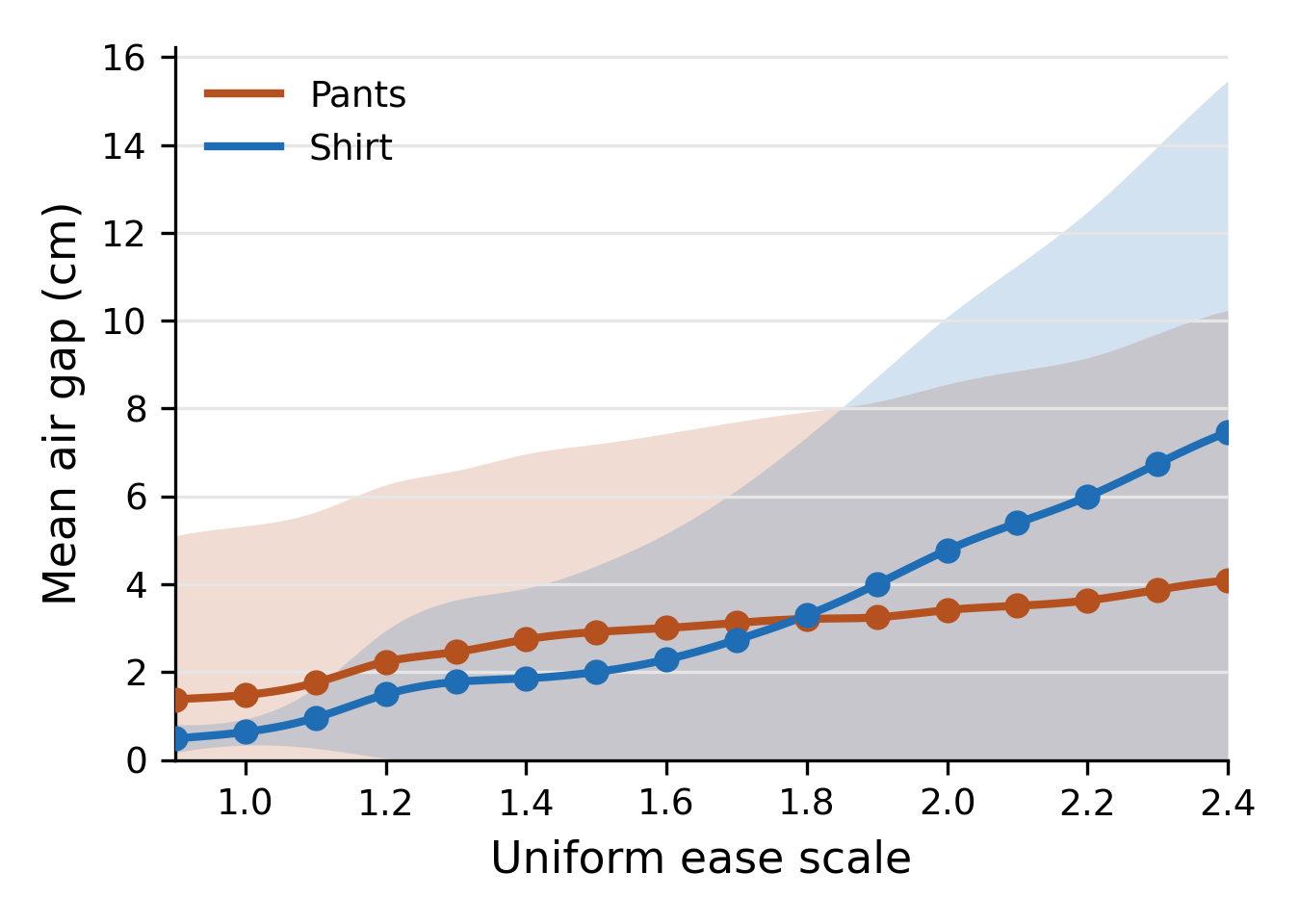}
    \vspace{-2.0em}
    \caption{Mean body-garment air gap as a function of uniform ease scale, for a shirt (upper) and pants (lower). Shaded regions indicate $\pm$1 standard deviation across body vertices.}
    \label{fig:gap-graph}
\end{figure}

To verify that the prescribed ease faithfully translates to physical fit, we measure the body-garment air gap after draping. For each vertex of the embedded garment patches, a ray is cast along the outward vertex normal; the gap at that vertex is the distance to the first intersection with the draped garment mesh. Vertices whose ray does not intersect the garment (e.g., near openings or due to penetration) are assigned zero gap. This ray-based formulation correctly handles tubular garments such as pants, where a nearest-surface distance would conflate the two legs. For each of 17 uniform ease levels between $\rho = 0.9$ and $\rho = 2.4$, we optimize a sewing pattern, drape it using the XPBD solver, and compute the per-vertex gap. Fig.~\ref{fig:gap-graph} shows the mean gap with standard deviation for a shirt (upper) and pants (lower). Both garments exhibit a monotonic increase in mean gap across the full ease range. The standard deviation grows at higher ease values, reflecting increased spatial variability as the garment departs further from the body surface.

\subsection{Garment Manufacturing and Body Measurement}

\begin{figure*}[t]
    \centering
    \includegraphics[width=.9\linewidth]{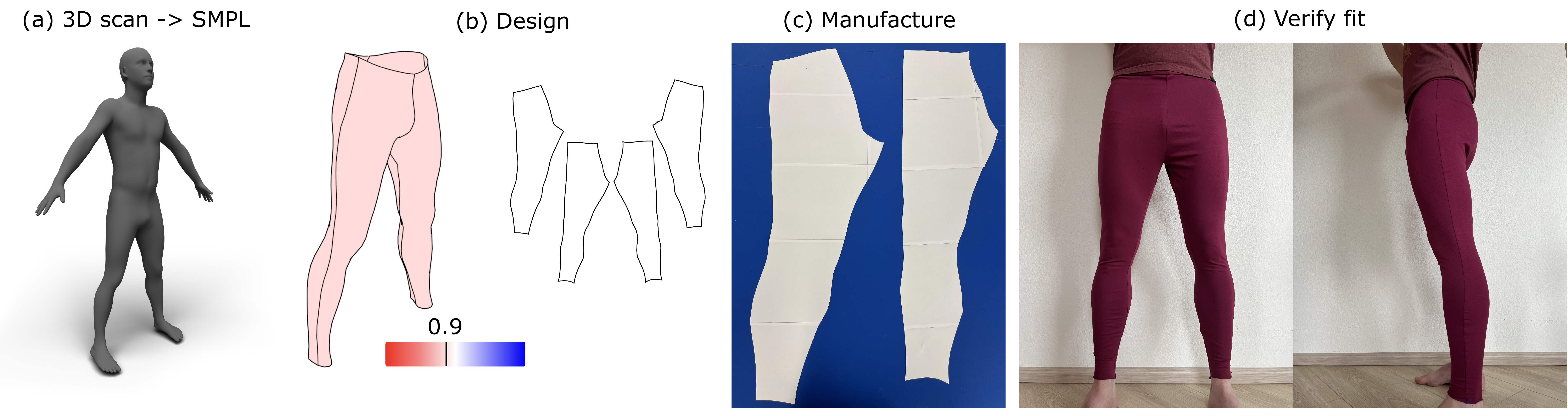}
    \vspace{-1em}
    \caption{Tight pant manufacturing sample and fit (negative ease). The subject is 3D scanned, the SMPL model is fitted to the scan, the uniform ease is specified, and the framework produces the sewing pattern. The pattern is physically realized and worn, verifying the intended tight fit.}
    \label{fig:pant-manufacturing}
\end{figure*}

To validate our sewing patterns in a real-world setting, we fabricated a pair of pants, shown in Fig.~\ref{fig:pant-manufacturing}. The subject is 3D scanned, and an SMPL mesh is fitted to the scan by optimizing $\boldsymbol{\beta}$ to match a small set of equidistant torso and leg circumferences anchored at the belly button landmark. This procedure is sufficient for lower-body garment design and avoids extracting full standard body measurements from scans~\cite{a-review-of-body-measurement, iso:7250}.
For manufacturing, we use a cotton fabric with less than 5\% stretch along the grain directions (Fig.~\ref{fig:pant-manufacturing}). The intended design is tight-fitting, corresponding to 95\% of the body circumference.

\begin{table}[t]
\centering
\caption{\revisedtag\ Ease compliance. Normalized deviation between specified and achieved ease, aggregated across upper and lower garments. Top: uniform ease ranges. Bottom: non-uniform designs.}
\label{tab:ease-compliance}
\begin{tabular}{l r}
\toprule
Range or design & Mean $\pm$ Std ↓ (\%) \\
\midrule
\multicolumn{2}{l}{\textit{Uniform}} \\
Tight (0.9--1.1)   & 1.3 $\pm$ 1.7 \\
Regular (1.2--1.5) & 1.7 $\pm$ 2.0 \\
Loose (1.6--2.0)   & 2.2 $\pm$ 2.5 \\
\midrule
\multicolumn{2}{l}{\textit{Non-uniform}} \\
Teaser shirt (Fig.~\ref{fig:teaser})       & 1.2 $\pm$ 1.6 \\
Simple dress, linear (Fig.~\ref{fig:analyses})     & 2.1 $\pm$ 2.7 \\
\bottomrule
\end{tabular}
\end{table}

\begin{table}[t]
\centering
\caption{Seam compatibility and performance. Seam mismatch measures normalized seam length differences between corresponding pattern edges.}
\label{tab:seam-runtime}
\begin{tabular}{l r r}
\toprule
Garment & Mean $\pm$ Std ↓ (\%) & Runtime ↓ (s) \\
\midrule
Upper & 3.7 $\pm$ 2.6 & 13.6 $\pm$ 0.1 \\
Lower & 3.1 $\pm$ 3.0 & 14.2 $\pm$ 0.1 \\
\bottomrule
\end{tabular}
\end{table}

\begin{table}[t]
\centering
\caption{Design transfer deviation. Normalized per-triangle ease deviation after transferring the initial design to four target body shapes, aggregated across upper and lower garments.}
\label{tab:transfer}
\begin{tabular}{l r r r r}
\toprule
 & XS & S & L & XL \\
\midrule
Mean ↓ (\%) & 1.0 & 0.6 & 1.0 & 1.8 \\
\bottomrule
\end{tabular}
\end{table}

\begin{table}[t]
\centering
\caption{Residual stretch for design adaptation. Percentage of garment surface area requiring stretch beyond material tolerance, before and after adaptation, per target pose and garment type.}
\label{tab:residual-stretch}
\begin{tabular}{l l r r}
\toprule
Garment & Pose & Before (\%) & After ↓ (\%) \\
\midrule
Upper & P1 & 28.0 & 4.5 \\
      & P2 & 27.2 & 4.3 \\
      & P3 & 34.3 & 4.4 \\
\midrule
Lower & P1 & 1.5 & 0.5 \\
      & P2 & 1.3 & 0.3 \\
      & P3 & 9.9 & 2.5 \\
\bottomrule
\end{tabular}
\end{table}

\begin{figure*}[t]
    \centering
    \includegraphics[width=.95\linewidth]{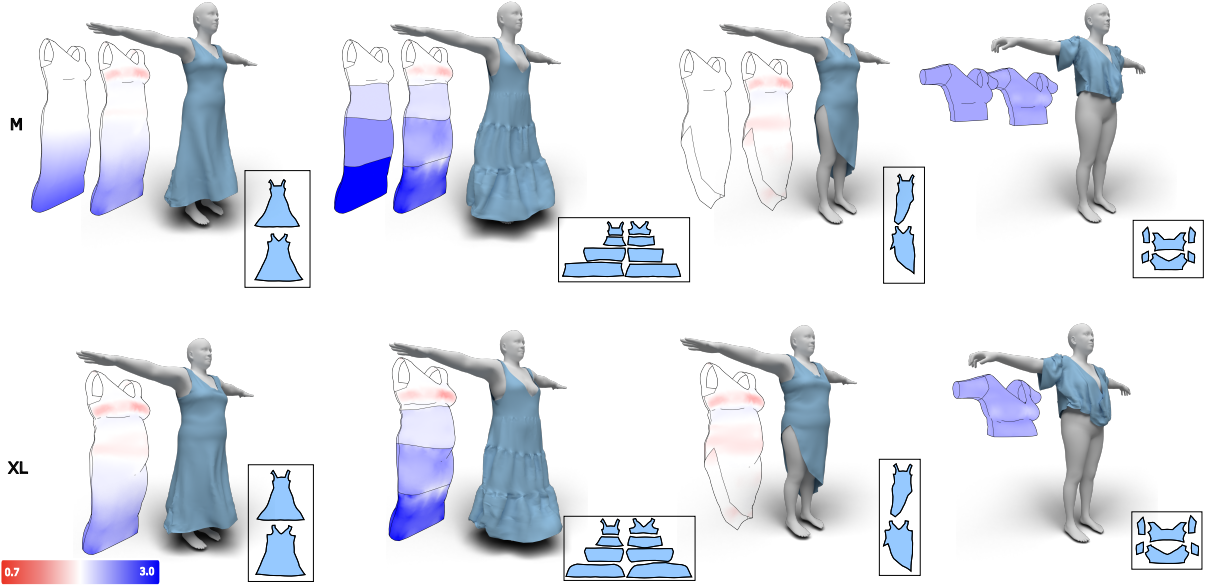}
    \vspace{-1em}
    \caption{Analyses of ease distribution and consistency across four representative designs, highlighting specified and optimized ease on the reference and target shape. The creation of designs in our user interface takes between 5--15 minutes. The garments are optimized and draped in the proposed framework.}
    \label{fig:analyses}
\end{figure*}

\begin{figure}
    \centering
    \includegraphics[width=\linewidth]{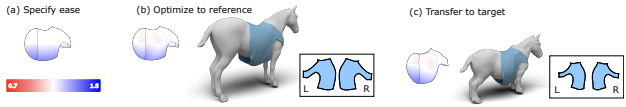}
    \vspace{-2.5em}
    \caption{The versatility experiment. We design a simple 2-piece garment for horses using the VAREN model and transfer it to a target shape.}
    \label{fig:horse-experiment}
\end{figure}
\section{Discussion}
\label{sec:discussion}

\textbf{Intent-preserving design.} The ease in the female bust area is typically slightly lower than specified, which is particularly visible in Fig.~\ref{fig:analyses}. To further improve the compliance to the specified intent, a trivial solution is to increase the specified ease in the bust region, but a more principled solution requires additional cuts such as darts. The design transfer is consistent across body shapes, as demonstrated qualitatively and quantitatively.

\textbf{Design adaptation.} When we tried to drape the unadapted garment in the target pose in Fig.~\ref{fig:teaser}, the simulation failed due to significant fabric stretches around the shoulder area. The final simulation shows a successful drape after the adaptation. As expected, the adapted sewing pattern (c) is asymmetric. In practice, the design can also be jointly adapted to the opposite side using mirrored pose. The max-based formulation explicitly ensures, by construction, that the optimized 2D pattern is at least as large as the body requires in each target pose — the fabric is not required to stretch beyond $\epsilon_{\text{weft}}$ to cover the body in any pose. While conservative, this provides a feasible solution without solving a more complex constrained optimization, and the stretch tolerance $\epsilon_{\text{weft}}$ offers a principled handle for relaxing this when fabric elasticity permits. The resulting surplus fabric in the reference pose can be reduced by increasing the stretch tolerance (last row in Fig.~\ref{fig:qualitative}), allowing the fabric to stretch locally rather than adding material. Therefore, the proposed design adaptation strategy can be applied for physical garment design or prototyping, but also to efficiently resolve possible collisions on animated virtual characters.

\begin{figure}
    \centering
    \includegraphics[width=\linewidth]{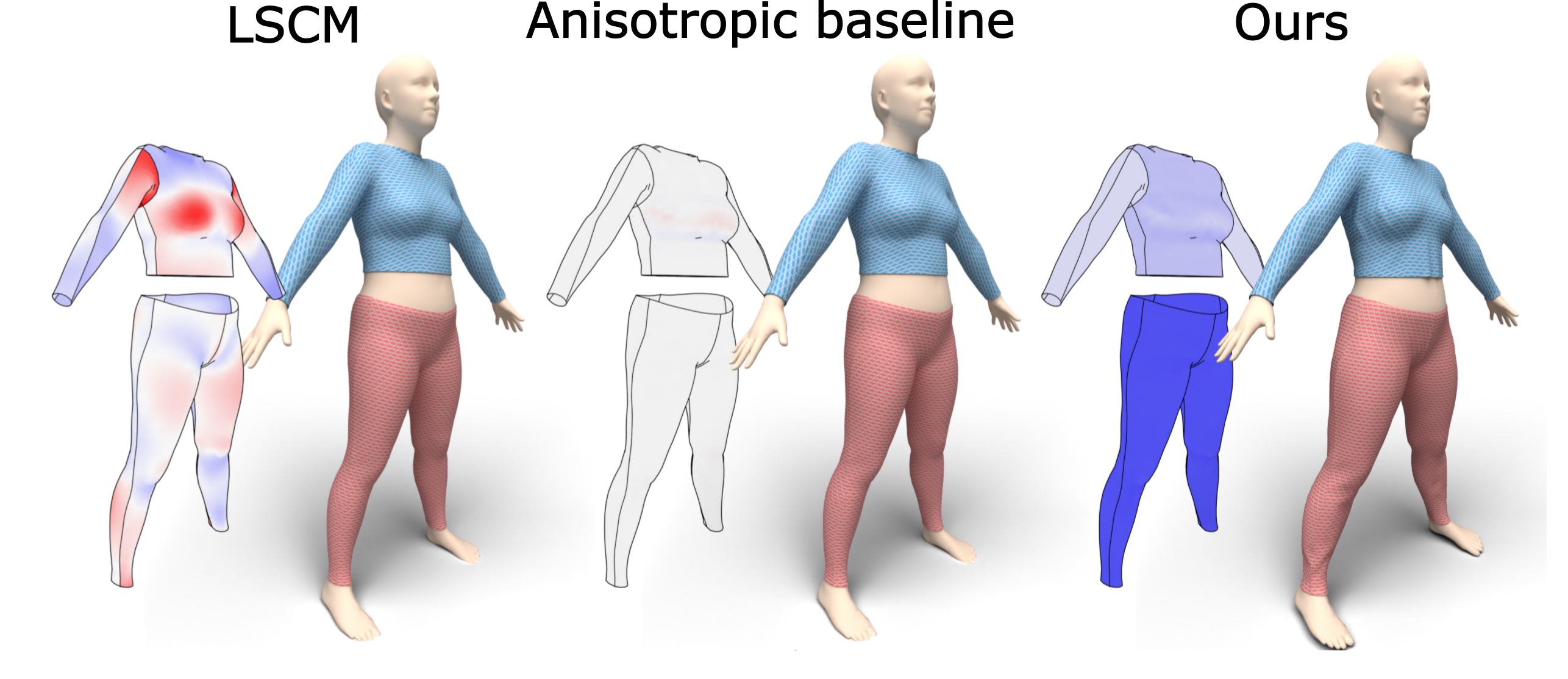}
    \vspace{-1.5em}
    \caption{\newtag\ Comparison with LSCM and Pietroni et al.~\cite{computational-pattern-making} on a skintight shirt and pants. Color encodes per-triangle weft scale.}
    \label{fig:baselines}
\end{figure}

\begin{figure}
    \centering
    \includegraphics[width=\linewidth]{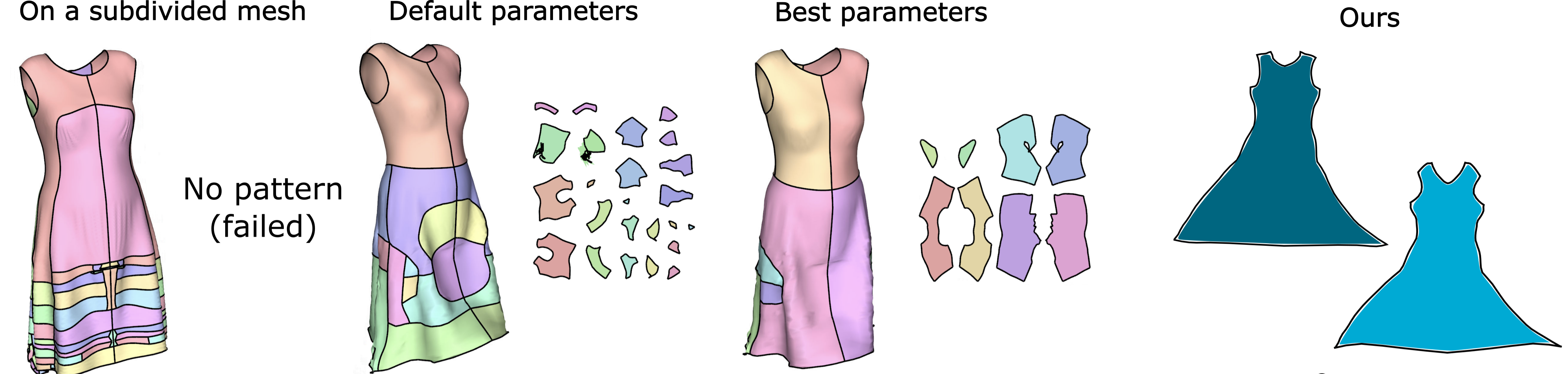}
    \vspace{-1.5em}
    \caption{\newtag\ Comparison with Pietroni et al.~\cite{computational-pattern-making} on a dress with non-uniform ease (tighter bust, broader hem).}
    \label{fig:pietroni-vs}
\end{figure}

\begin{newpar}
    \textbf{Comparison with baselines.} On a skintight shirt and pants design, we compare our method to two baselines that share input geometry: a standard LSCM parameterization~\cite{lscm}, and the anisotropic-aware baseline ~\cite{computational-pattern-making}, which is the closest to our formulation (Fig.~\ref{fig:baselines}). LSCM produces visible weft stretch in the bust region ($\rho \approx 0.7$). The anisotropic baseline reduces this stretch to near-unit scale ($\rho \approx 1.0$), closely corresponding to the intended skintight design, but does not expose ease as a design variable. Our method specifies a uniform per-triangle ease scale field, producing a loose garment ($\rho \approx 1.2$). Furthermore, we compare against ~/cite{computational-pattern-making} on a dress with non-uniform ease (tighter bust, broader hem) in Fig.~\ref{fig:pietroni-vs}. Across the three configurations (subdivision, default, best), the method either fails to produce a pattern or produces fragmented panels with irregular cut-outs that are not manufacturable. Our method produces two compatible panels suitable for sewing. On the other hand, \cite{computational-pattern-making} cuts patches from scratch without our predefined seamlines, which is a structurally harder setup. Still, the result shows that decomposing ease specification from seamline positioning significantly simplifies the optimization and improves the final result.
\end{newpar}

\begin{wrapfigure}[14]{r}{0.22\linewidth}
    \centering
    \vspace{-1em}
    \includegraphics[width=\linewidth]{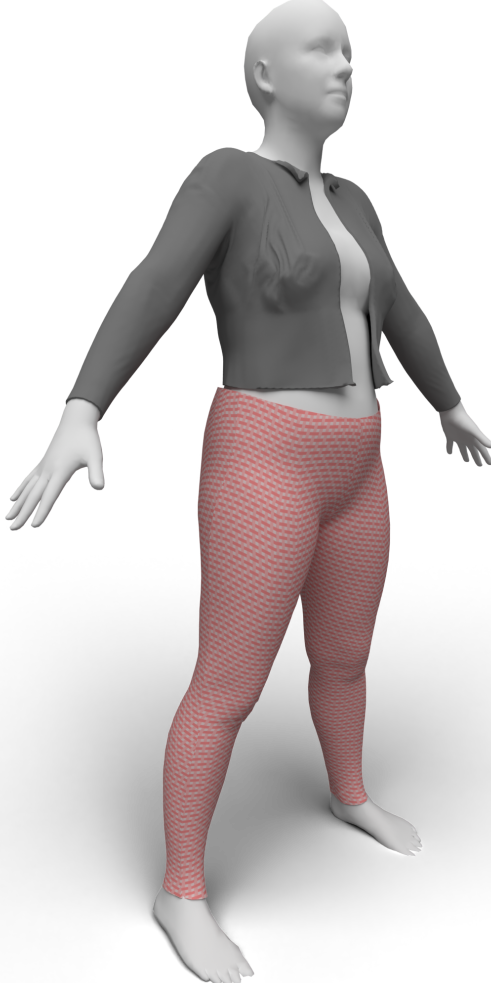}
    \vspace{-1.5em}
    \caption{\newtag\ A jacket is made by cutting the front of the shirt.}
    \label{fig:jacket}
\end{wrapfigure}
\textbf{Post-drape gap.} The monotonic correspondence between prescribed ease and measured air gap (Fig.~\ref{fig:gap-graph}) confirms that the ease field reliably controls physical material allowance through the full pipeline. The absolute gap values depend on the simulator's material parameters and collision response, which are held fixed across all experiments; the relative ordering, however, is robust to these choices as it is determined by the pattern geometry rather than the material model.

\textbf{Design expressivity.} Fig.~\ref{fig:analyses} highlights four representative designs. The first design is a standard long dress. Instead of linearly increasing looseness from top to bottom, the ease is typically achieved using gathers. The second design shows a dress with four components, each having a significantly increased body-relative uniform ease (up to 3x the corresponding skirtified body surface). The third design highlights cutting expressivity, where the dress surface used in the first two examples is cut to open the right leg. The final design shows a uniformly loose top (80\% additional ease). This also highlights the limitation of our representation, where garments that more strongly depart from the body surface are more challenging to model. The same cutting mechanism extends to a different garment family in Fig.~\ref{fig:jacket}, where a cut along the front center transforms the long-sleeve upper into a jacket.

\textbf{Horse experiment.} To demonstrate the versatility of our approach to other parametric models, we design a garment for the horse (Fig.~\ref{fig:horse-experiment}) using the VAREN model \cite{varen}. The design specifies linearly increasing ease in the chest area, up to 30\% at the bottom. The optimized design closely matches the specified ease, while the sewing pattern seamlines are compatible. The design is transferred to a significantly different, target horse shape, smaller and with a shorter torso. The design intent is preserved, as indicated by the matching ease distribution and the final drape.

\textbf{Execution time.} The cutting and patch extraction takes around 5 seconds. The parameterization algorithm takes 1-2 seconds on a CPU, depending on the number of target poses. The cloth simulation using XPBD solver in Newton takes around 5 seconds on a GPU and around 30-45 seconds on a CPU. In total, it takes around 10-15s on GPU (35-50s on CPU) to cut the mesh, extract the patches, optimize the initial and several target poses, and drape the result. Although our method is not yet optimized for efficiency, it is already among the fastest approaches for comparable tasks in prior work, e.g., compared to Wolff et al. \cite{wolff} ($\sim$2.8min for two target poses), Wang \cite{rule-free-pattern-adjustment} (few minutes), DressAnyone \cite{dress-anyone} (5-30min per target body shape), and DiffAvatar \cite{diffavatar} (20-200min).

\textbf{Implementation details.} Draping is performed using Newton’s XPBD solver, initialized directly from the optimized sewing patterns. The solver derives anisotropic rest lengths and stretch constraints from the 2D–3D correspondences produced by the parameterization stage, and resolves gravity and body collisions under a unified constraint formulation. Simulation is used solely to visualize the final draped configuration and does not influence the optimized ease distribution. The project is developed and tested under MacOS on an Apple M3 Pro (CPU-only) using Python and C++, and on a PC under Linux (GPU). Rendering is done separately using Blender Cycles \cite{blender}. The default optimization parameters are ($\lambda_{\text{ease}} = 0.1$, $\lambda_{\text{reg}} = 1$, $\lambda_{\text{len}}=100$, $\lambda_{\text{straight}}=5$, $\lambda_{\text{curve}}=10$). When using gathers (Fig.~\ref{fig:analyses}), we turn off the seam length compatibility term.
\section{Conclusion}

We presented a garment design framework that represents ease explicitly as spatially varying, anisotropic scale fields embedded on a parametric human body model. By decoupling the specification of material allowance from physical simulation,
our approach enables efficient transfer and adaptation of garment designs across body shapes and poses directly in the design space. This formulation complements recent simulation- and learning-based approaches to garment adaptation by providing explicit, editable control over local fit, while remaining compatible with downstream physics-based draping. The manufacturing experiment verifies that the resulting patterns can be realized using standard cut-and-sew workflows for woven fabrics.

\textbf{Limitations.}
Although the design intent is realized on the reference body and preserved between the shapes, a perfect fit requires the definition of darts, especially in female bust area, which we leave for future work. The material stretch tolerance is implemented, but could be made more sophisticated by applying it along the entire fabric grain rather than locally. Highly silhouette-driven styles, such as boxy oversized garments, pleated or gathered designs, or garments with volumetric shaping that departs strongly from the body, may require additional design variables, such as topology changes, explicit fold modeling, or volumetric constraints, which are not yet captured by our representation.
Bijectivity of the parameterization is not explicitly enforced; we did not observe triangle flips in any reported experiment, though degenerate configurations with extreme ease gradients may require increasing $\lambda_{\text{reg}}$.

\section*{Acknowledgements}

This work is partially funded by the German Research Foundation (DFG) as part of EXC 2050/2, Project ID 390696704, Cluster of Excellence ``Centre for Tactile Internet with Human-in-the-Loop'' (CeTI) of TU Dresden; by the German Federal Ministry of Research, Technology, and Space (BMFTR) as part of project 6G-life (16KIS2414); by DFG grant 389792660 as part of TRR 248 (CPEC); and by the German Federal Ministry of Education and Research (BMBF, SCADS22B) and the Saxon State Ministry for Science, Culture and Tourism (SMWK) by funding the Center for Scalable Data Analytics and AI ``ScaDS.AI Dresden/Leipzig''.

{
    \small
    \bibliographystyle{ieeenat_fullname}
    \bibliography{refs}
}

\end{document}